# Chip-based frequency combs sources for optical coherence tomography

Xingchen Ji,[1,2] Alexander Klenner,[3] Xinwen Yao,[1] Yu Gan,[1] Alexander L. Gaeta,[3] Christine P. Hendon[1] and Michal Lipson[1,*]

[1]Department of Electrical Engineering, Columbia University, New York, NY 10027
[2]School of Electrical and Computer Engineering, Cornell University, Ithaca, NY 14853
[3]Department of Applied Physics and Applied Mathematics, Columbia University, New York, NY 10027
*ml3745@columbia.edu

**Abstract**

The Optical coherence tomography (OCT) is a powerful interferometric imaging technique widely used in medical fields such as ophthalmology, cardiology and dermatology, for which footprint and cost are becoming increasingly important. Here we present a platform for miniaturized sources for OCT based on chip-scale lithographically-defined microresonators. We show that the proposed platform is compatible with standard commercial spectral domain (SD) OCT systems and enable imaging of human tissue with an image quality comparable to the one achieved with tabletop commercial sources. This platform provides a path towards fully integrated OCT systems.

Optical coherence tomography (OCT) is a non-invasive imaging modality that provides depth-resolved, high-resolution images of tissue microstructures in real-time. OCT has been widely demonstrated in medical fields such as ophthalmology and cardiology[1]. Recently, great efforts have been spent on the development of on-chip OCT components in order to enable OCT systems with small footprint and cost[2–6]. These efforts have leveraged recent advances in photonic integration on-chip including the beam splitter, reference arm, sampling arm and spectrometer[2–6]. However, the degree of miniaturization of the OCT system based on the miniaturization of these components is limited, since these systems still rely on an external, tabletop light source such as a superluminescent diode (SLD) or swept source laser that cannot be easily integrated with current photonics on a silicon platform.

Here we introduce a platform for a miniaturized OCT source based on chip-scale lithographically-defined microresonators. These microresonators are fabricated using traditional microelectronic processes. When optically pumped with a single continuous-wave laser source they can generate broadband frequency combs, consisting of discrete lines with a frequency spacing determined by the geometry of the resonator. Such frequency combs have been demonstrated in numerous chip-scale platforms including silica[7–10], silicon[11,12], silicon nitride[13–15], aluminum nitride[16], crystalline fluorides[17,18], diamond[19] and AlGaAs[20] in the past decade.

Our platform is based on ultra-low loss silicon nitride resonator generated frequency combs[13–15,21]. This resonator platform when integrated with semiconductor amplifiers, has recently been shown to enable highly efficient broadband frequency comb generation on-chip[22]. Silicon nitride ($Si_3N_4$) combines the beneficial properties of a wide transparency range covering the entire OCT imaging window, a high nonlinear refractive index ($n_2 = 2.4 \times 10^{-19}$ m²/W),[23] and semiconductor mass manufacturing compatibility. $Si_3N_4$ frequency combs have recently been recently generated using a reflective semiconductor amplifier as a pump source enabling a millimeter sized electrically pumped source[22]. Fig 1 shows an artist's view of a chip-scale OCT system consisting of lithographically defined components all on a single chip, where the $Si_3N_4$ microresonator acts as a light source. Fig 1 inset shows the recently demonstrated hybrid approach for achieving chip-scale mm-size electrically pumped microresonator combs[22].

In order to enable a large imaging range using the optical comb as an OCT source of at least 2 mm (comparable with commercial OCT imaging range), we design the combs with a small spectral line spacing of 0.21 nm (corresponding to 38 GHz) using a large microresonator with a perimeter of 1.9 mm (see the Supplementary Materials). This perimeter is at least an order of magnitude larger than traditional high confinement micro-resonators[14,24,25]. In order to achieve sufficient optical power build up and enable comb generation in such a large cavity[8,26], we rely on the extremely low loss platform recently demonstrated[27]. The ultra-low loss of 3 dB/m compensates for the large mode volume and enables frequency combs generation with 120 nm bandwidth. In order to generate the combs with broad bandwidth and high conversion efficiency, ideal for OCT imaging, we ensure that the combs generation process does not induce soliton states with characteristic hyperbolic secant spectrum, by tuning of the cavity resonance relative to the pump frequency using a microheater co-fabricated with the resonator[28–30]. Figure 2A shows the fabricated on-chip resonator. Figure 2B shows the generated frequency comb spectra using a ring resonator based on waveguides with 730×1500 nm cross section. The measured power in these frequency combs lines is 42 mW with pump power of 142 mW corresponding to 30% conversion efficiency.

Using the microresonator platform, we acquire OCT images of human tissue with chip-based frequency combs and show that the platform is compatible with a standard commercial SD-OCT system[31]. These images were achieved using a standard SD-OCT system (Thorlabs Telesto I), where the SLD was simply replaced by the chip-based frequency combs. Since the system is not optimized for our combs, the imaging capability is a lower bound limit. Figures 3-4 show *ex vivo* OCT images of human breast and coronary artery samples imaged with our microresonator frequency comb source using a commercial SD-OCT system[31]. The human breast tissue was obtained from Columbia University Tissue Bank[32], and the human heart was obtained via the national disease research interchange[33]. Figure 3 compares images recorded using our microresonator frequency comb and a commercial SLD which has similar performance to the generated combs (see the Supplementary Materials). The Hematoxylin and Eosin (H&E) stained histology is provided as the reference for both the breast and two arteries in cardiovascular system, coronary artery and aorta. Different tissue types, including stromal tissue, adipose tissue and milk duct are delineated in both B-scans by comparing with the corresponding histology analysis. Figure 4A shows a stitched frequency-comb-based OCT image of a human left anterior descending artery (LAD) in comparison with the H&E histology in Figure 4B. Figure 4C shows a stitched frequency-comb-based OCT image of a human aorta in comparison with the H&E histology in Figure 4D. OCT B-scans were stitched using the method previously used in cervical imaging[31]. In the red inset, a gradually decreasing trend of backscattering can be visualized within the transition region from a fibrous region to the media. The blue inset in Figure 4 reveals a typical pattern of a fibrocalcific plaque[3], where a layer of signal-rich fibrous cap is on the top of calcium, a signal-poor region with a sharply delineated border. Importantly, overlying the fibrocalcific plaque region, we can see a transition from dense fibrous cap a region with a thinner fibrous cap for unstable plaque structure. The green inset in Figure 4 shows the visualization of large calcification region, the deposit of calcium. Figures 3 and 4 show the potential to visualize critical features within human breast and cardiovascular samples by integrating the chip-based frequency combs into an OCT system.

We have demonstrated the viability of chip-based frequency comb platform as light sources for OCT systems a key step toward fully integrated chip-scale OCT systems. The different building blocks needed in order to realize an integrated OCT system including a chip-scale beam splitter, reference arm, sampling arm and spectrometer have already been demonstrated recently and can be integrated on the same chip as the microresonator[2–6] enabling the miniaturization and lower cost of OCT systems.

In addition to enabling highly integrated sources, $Si_3N_4$ microresonator combs exhibit a bandwidth that is determined by the waveguide geometry alone and not limited by the optical power[13,30], in contrast to traditional OCT sources based on SLD sources with limited bandwidth at high optical powers due to gain narrowing. With waveguide dispersion engineering and a spectrometer designed for the combs, this platform could enable high axial resolution and high penetration depth.

**Methods**

Device fabrication

Starting from a silicon wafer, a 4-µm-thick oxide layer is grown for the bottom cladding. Silicon nitride ($Si_3N_4$) is deposited using low-pressure chemical vapor deposition (LPCVD) in two steps. After $Si_3N_4$ deposition, we deposit a silicon dioxide ($SiO_2$) hard mask using plasma enhanced chemical vapor deposition (PECVD). We pattern our devices with JEOL 9500 electron beam lithography. Ma-N 2403 electron-beam resist is used to write the pattern, and the nitride film is etched in an inductively coupled plasma reactive ion etcher (ICP RIE) using a combination of $CHF_3$, $N_2$, and $O_2$ gases. After stripping the resist and oxide mask, we anneal the devices at 1200°C in an argon atmosphere for 3 hours to remove residual N-H bonds in the $Si_3N_4$ film. We clad the devices with 500 nm of high temperature silicon dioxide (HTO), deposited at 800°C, and followed by 2.5 µm of $SiO_2$ using PECVD. Chemical Mechanical Polishing (CMP) and multipass lithography technique can be applied to further reduce sidewall scattering losses[27]. Above the waveguide cladding, we fabricate integrated microheaters by sputtering platinum and using a lift-off approach. We integrated micro-heaters on our device to control the cavity resonance by temperature tuning, which enables the use of a simple compact single-frequency pump laser diode to generate frequency combs[15,34].

Measurements

As the presence of the pump within the comb spectrum limits the dynamic range of the detection, we use a filtering setup based on a free-space grating and pin to fully attenuate the pump power. The setup is shown in the

Supplementary Materials. This filtering setup can be replaced by a customized fiber-based filter or an on-chip filter to miniaturize the size of the setup in the future. We directly plug the comb source into a commercial system (Thorlabs Telesto I) to acquire images. The schematic of the OCT system is shown in the Supplementary Materials. An optical circulator with an isolation of -40dB is added to protect the commercial console. The incident light from the comb source is routed to the Michelson interferometer, and the backscattered signals from both interferometer arms are directed back to the spectrometer.

Using the frequency combs combined with the commercialized SD-OCT system, we are able to acquire OCT images. The images are reconstructed in real-time from the raw spectral data generated by the system, following standard OCT signal processing steps, including background subtraction, linear-k interpolation, apodization, and dispersion compensation. The acquisition rate is 28 kHz currently limited by the CCD line rate. The total acquisition time of an image for the SLD and the chip comb images is the same (35msec). The sensitivity of the OCT system is defined by the minimal sample reflectivity at which the signal to noise ratio reaches unity[35]. It is measured to be 98 dB at an A-line rate of 28 kHz with the frequency comb source. The sensitivity can be further increased by suppressing the noise due to the laser-chip coupling via packaging[36].

**Acknowledgements**

The authors would like to thank Charles Marboe for his histopathological assistance. This work was performed in part at the Cornell NanoScale Facility, a member of the National Nanotechnology Coordinated Infrastructure (NNCI), which is supported by the National Science Foundation (ECCS-1542081). The authors acknowledge support from the Defense Advanced Research Projects Agency (N66001-16-1-4052), the Air Force Office of Scientific Research (FA9550-15-1-0303), the National Science Foundation (2016-EP-2693-A, CCF-1640108) and the National Institute of Health (1DP2HL127776-01). X.J. acknowledges the China Scholarship Council for financial support.

**Author contributions**

X.J. prepared the manuscript in discussion with all authors. X.J., A.K. and X.Y. designed and performed the experiments. X.J. fabricated the devices. A.K. performed theoretical modelling and simulations. X.Y. and Y.G. performed the OCT measurements and data analysis. M.L., C.P.H. and A.L.G. supervised the project.

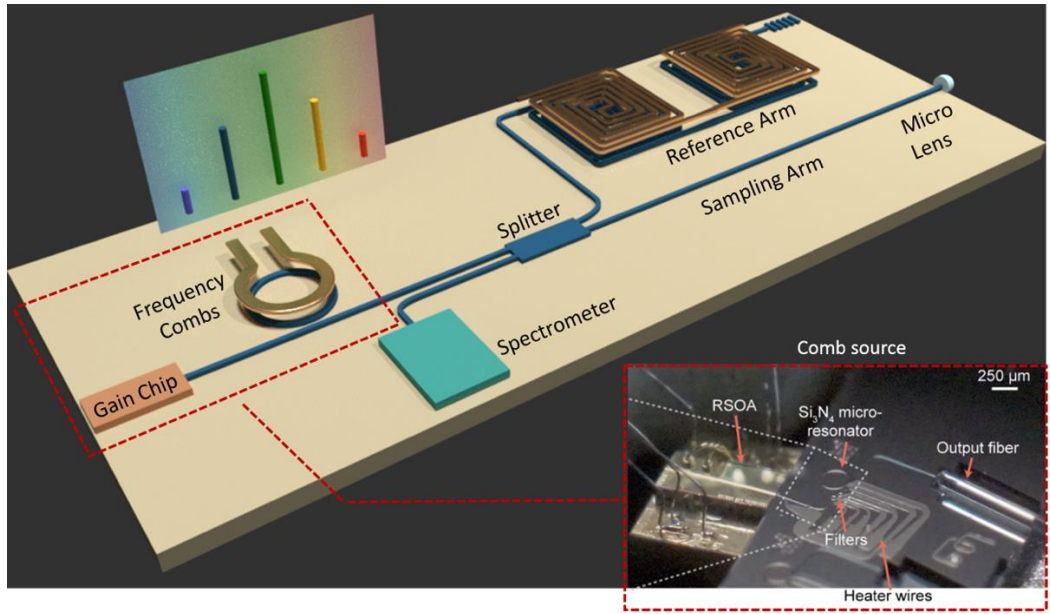

**Fig 1.** Artist view of a fully integrated OCT systems with frequency combs light source. The frequency combs light source is formed by a reflective semiconductor optical amplifier chip fully integrated with an ultra-low loss $Si_3N_4$ microresonator (Inset). The interferometer, including beam splitter, reference arm, sampling arm and spectrometer is integrated on the same chip. A mircolens and MEMS based scanning mirror can be attached to emit and collect and backscattered light from the sample[6].

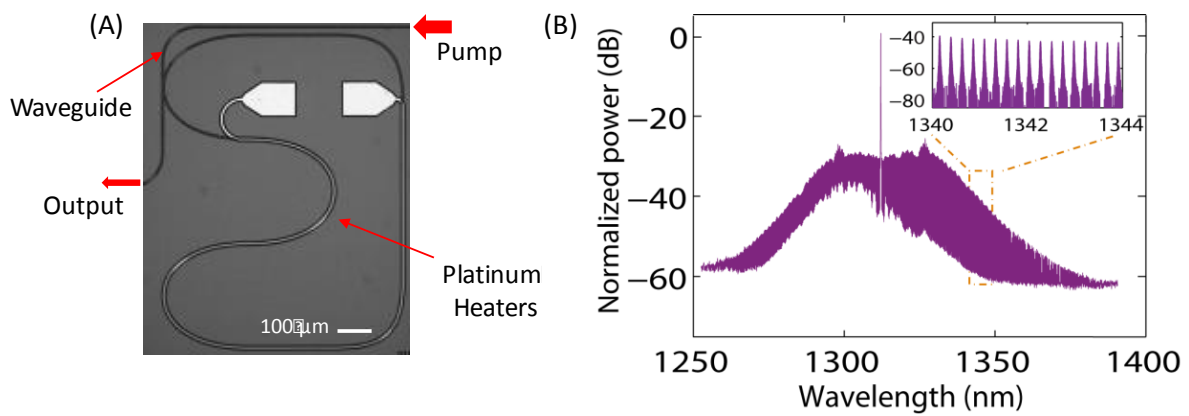

**Fig 2.** Device image and measured spectrum. (A) Microscopy image of the silicon nitride on-chip microresonator. A platinum heater is fabricated over a large portion of the cavity and allows electric contact via the pads. (B) Measured frequency comb spectrum generated using the silicon nitride microresonators. Inset shows line spacing of 0.21 nm.

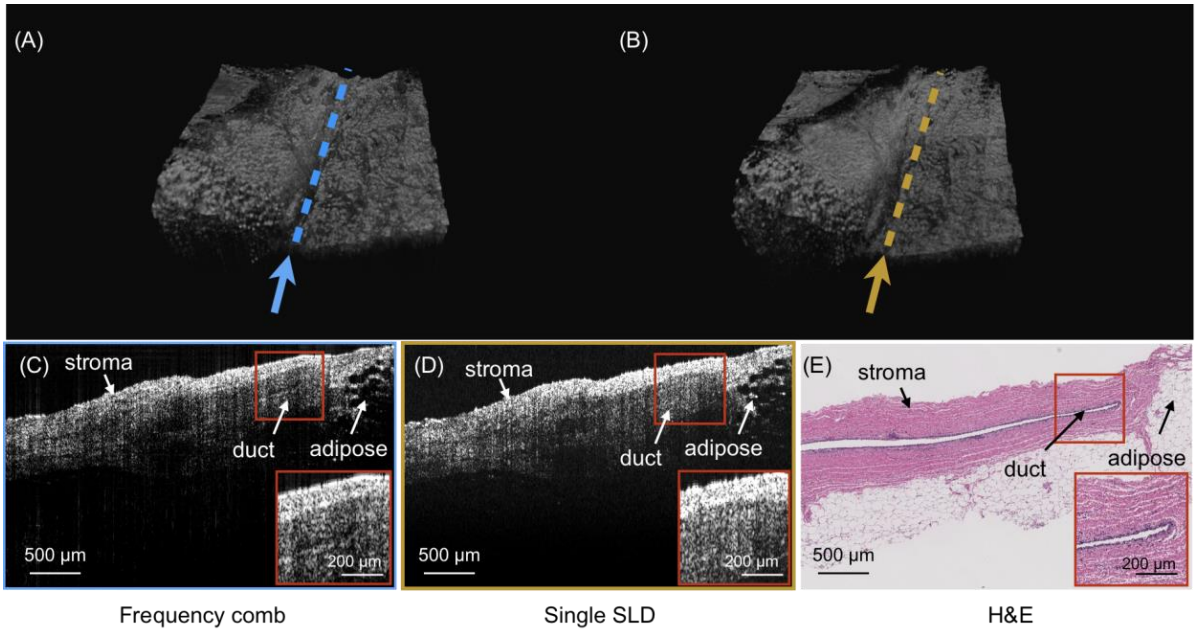

**Fig 3.** OCT images comparsion. OCT C-scans of human breast tissue taken with (A) the frequency comb source, (B) a single SLD source, and OCT B-scans of the same tissue taken with (C) the frequency comb source (marked by the blue arrow) and (D) a single SLD source (marked by the yellow arrow), respectively, corresponded with (E) the H&E staining slide. Different features and tissue types, such as stromal tissue, adipose tissue and milk duct, are delineated in both B-scans.

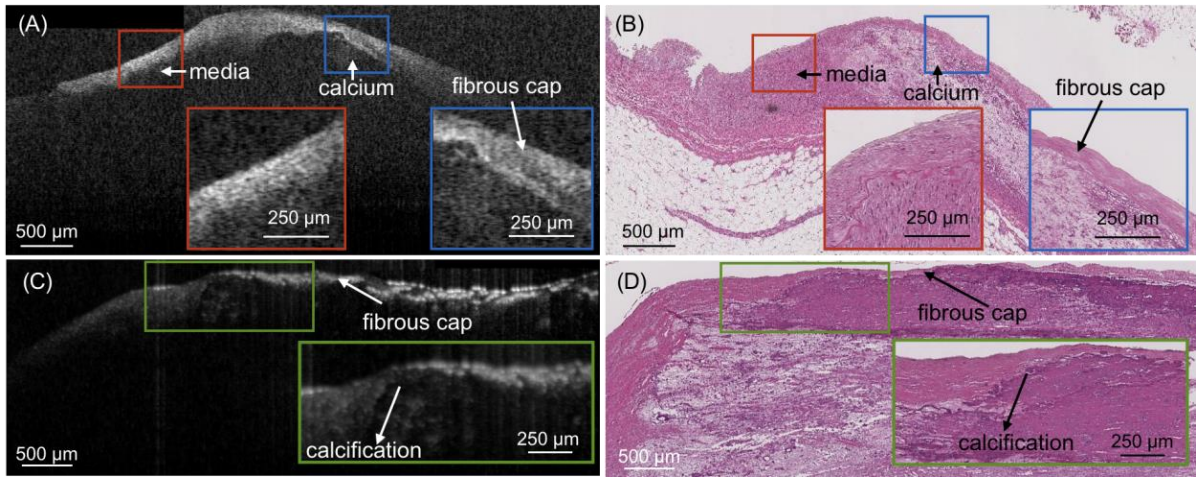

**Fig 4.** Frequency-comb-based OCT images. Stitched frequency-comb-based OCT B-scans of human coronary artery (A) and aorta (C) with corresponding H&E histology of coronary artery (B) and aorta (D). Critical features are observed, including delineation of the fibrous cap, calcium, and layered structure of intima and media are depicted within OCT images.